\documentclass[aps,pre,onecolumn]{revtex4}
\usepackage[dvips]{graphicx}
\usepackage{dcolumn}
\usepackage{bm}
\usepackage{amsmath,amssymb}
\usepackage{mathrsfs}
\usepackage[dvips]{color}
\usepackage{ulem}
\usepackage{makeidx}
\makeindex

\begin{document}

\title{Properties of the connected components in projections of random bipartite networks: \\ Effects of clique size fluctuations}

\author{Yuka Fujiki}\email{yuka.fujiki.a1@tohoku.ac.jp}
\affiliation{Frontier Research Institute for Interdisciplinary Sciences (FRIS), Tohoku University, 6-3 Aramaki aza Aoba, Aoba-ku, Sendai 980-8578, Japan}
\affiliation{Advanced Institute for Materials Research (AIMR), Tohoku University, 2-1-1 Katahira, Aoba-ku, Sendai, 980-8577 Japan}
\author{Shogo Mizutaka}\email{shogo.mizutaka.kh50@vc.ibaraki.ac.jp}
\affiliation{Graduate School of Science and Engineering, Ibaraki University, 4-12-1 Nakanarusawa, Hitachi, Ibaraki 316-8511, Japan}

\date{\today}
\begin{abstract}
We examined the structure of projections of random bipartite networks characterized by the degree distribution of individual and group nodes through the generating function method.
We decomposed a projection into two subgraphs, the giant component, and finite components and analyzed their degree correlation.
The projections never exhibit a negative degree correlation.
Positive degree correlations in the networks originating from the clique size fluctuation remain after the decomposition at the set of finite components although the values of their clustering coefficient are still finite.
The giant component can exhibit either positive or negative degree correlations based on the structure of the projection. However, they are positively correlated in most cases.
In addition, we determined the relation between the finite components in a supercritical phase possessing the giant component and the projection in a subcritical phase when the degree distributions for group and individual are Poisson.
\end{abstract}
\maketitle

\section{Introduction}
\begin{figure}[t]
\begin{center}
\includegraphics[width=0.35\textwidth]{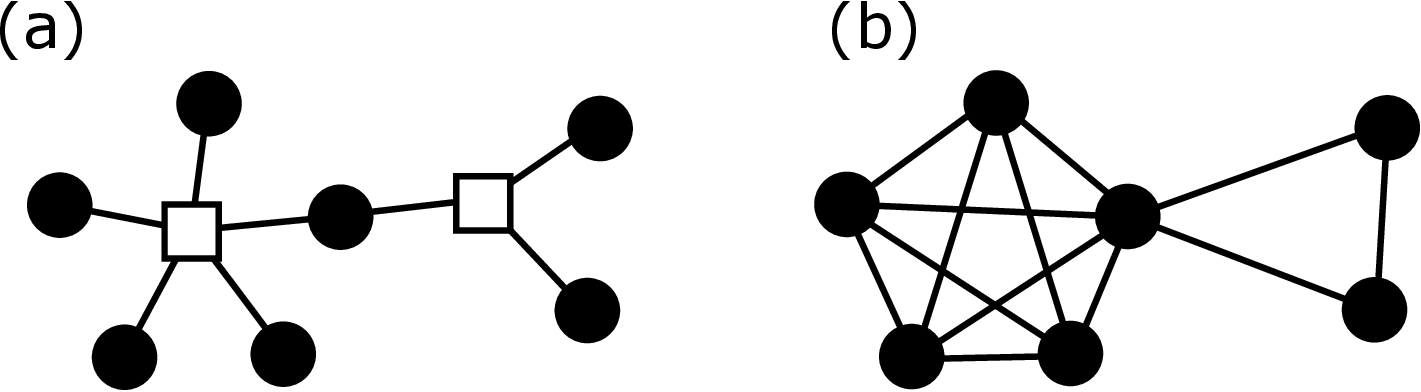}
\caption{
A schematic view of a bipartite network consisting of seven individual nodes (solid spheres) and two group nodes (open squares) (a), and its projection (b).
In the projection, two individual nodes sharing at least one adjacent group node in the bipartite networks are connected by an edge to each other.
}
\label{fig:fig1}
\end{center}
\end{figure}

The structural complexity of networks consisting of nodes and edges emerges from pairwise relations between two nodes and higher-order structure induced by group relations among more than two nodes.
In scientific fields, several researchers collaborate to produce results.
Many molecules may relate to a chemical reaction intravitally.
In an ecosystem, some species may compete or coexist.
Such group relations are widely observed in empirical networks ranging from nature to society such as neural, biological, ecological, social, and technological ones \cite{newman2002random,milo2002network,ugander2012structural,petri2014homological,benson2016higher,levine2017beyond,grilli2017higher,benson2018simplicial,sizemore2018cliques}.
Several studies focus on the higher-order structure by the language of pairwise networks.
One of the treatments is based on bipartite networks and their projections.
Bipartite networks consist of two disjoint sets of nodes and a set of edges between nodes of different sets, and they translate into projections by replacing a node set with cliques (Fig.~\ref{fig:fig1}).
A seminal study corresponds to the generating functions approach for bipartite networks \cite{newman2001random}.
Specifically, a random bipartite network characterized by two degree distributions of two node sets is utilized as a null model of a network with group structure.
The generating functions method provides clustering and assortativity coefficients for projections analytically and conjectures high levels of clustering and assortative mixing in social networks \cite{newman2001random,newman2002random,newman2003social}.
Detailed structural properties have been investigated for projections of random bipartite networks and empirical ones \cite{fisher2017perceived,vasques2018degree,vasques2020transitivity}.

Another important property is the emergence of the giant (bipartite) component, that is, the percolation transition.
A natural extension of the Molly--Reed criterion for the configuration model gives the transition condition \cite{newman2001random}.
With respect to a random (monopartite) network generated from Erd\H{o}s--R\'{e}nyi model and the configuration model, various structural properties of the giant component (GC) and set of finite components (SFC), which is the set difference of the whole network and the GC, are clarified so far.
A broad class of random networks satisfies a discrete duality relation; the SFC in a supercritical phase possessing the GC behaves like a random network in a subcritical phase possessing no GC \cite{bollobas2007phase,durrett2007random,janson2011duality}.
In the SFC, the degree distribution is characterized by that of the whole network with an exponential cutoff. Additionally, similar to the whole network, the degree correlation of the SFC is absent \cite{tishby2018revealing}.
The structural properties of the GC are different from the whole network and SFC. Specifically, the degree distribution of the GC coincides with the neighbor degree distribution of the whole network at the percolation threshold \cite{dorogovtsev2008critical}.
In terms of degree correlation, the GC is negatively correlated irrespective of the degree distribution of the whole network \cite{engel2004large,bialas2008correlations,tishby2018revealing,mizutaka2018disassortativity}, and the negative degree correlation extends to within the percolation correlation length \cite{mizutaka2020emergence}.
A few studies examine the properties of connected components in networks with group structure in the context of clustered network models.
The random clustered network model generates random networks in which independent distributions of single edges and triangles specify the degree of nodes \cite{newman2009random,miller2009percolation}.
The GC of a random clustered network can exhibit positive or negative degree--degree correlation based on the details of the clustering \cite{hasegawa2020structure}.
The generalized configuration model is a generalization of the random clustered network model, in which generated networks are specified arbitrary distributions of subgraphs \cite{karrer2010random}.
The generating function method for the generalized configuration model was used by Mann et al.\ to examine the structure of GC in a random network with arbitrary clique clustering \cite{mann2022degree}.
They observed that the GC possesses negative degree correlations for a single-size clique network.
They also investigated networks comprising single edges and $m$-cliques and noted that the average degree of adjacent nodes of a node increases with the node degree with an oscillation, implying a positive degree correlation.
Component statistics in random (monopartite) networks have been collected to a certain extent. However, a paucity of understanding of networks with group structure remains. In particular, it is not still completely understood what determines the degree correlation of components and how the structure in the subcritical and supercritical phases is related.

In the study, we treat networks with clique clustering that are projections of random bipartite networks \cite{newman2001random}.
In the projecting procedure, each group node with degree $m$ is replaced with a clique structure with size $m$ (Fig.~\ref{fig:fig1}).
The generating function method is used to analyze the statistical properties of GC and SFC of the projections of random bipartite networks and evaluate the assortativity \cite{newman2002assortativity} and global clustering coefficient \cite{newman2001globalclustering,barrat2000globalclustering}.
Our findings reveal that when all groups in a bipartite network possess the same size, the projection and the SFC are degree uncorrelated while the GC consistently exhibits a negative degree correlation irrespective of the individual node degree distribution.
Conversely, when employing the Poisson distribution for both degree distributions of individual and group nodes, GCs are positively degree correlated, except when the average degree of group nodes is sufficiently small.
We discuss the relation between the fluctuation of clique size and degree correlation in a network with tunable clique size fluctuation and show that a small amount of clique size fluctuation makes degree correlations in GC positive.

The remainder of this paper is organized as follows.
We formulate structural properties for projections and their subgraphs by using generating function methods in Sec.~\ref{sec:rh}.
We treat two examples to determine some properties of projections of bipartite networks in Sec.~\ref{sec:ex}.
Section~\ref{sec:end} presents conclusions and discussions.

\section{Formulation}\label{sec:rh}
We treat projections of random bipartite networks that consist of individual and group nodes.
In random bipartite networks, individual and group nodes are connected randomly.
The random bipartite networks in the study are locally tree-like and sparse.
In the next subsection, we briefly review the structural properties of whole network of projections of random bipartite networks \cite{newman2001random,newman2003social}.
In Sec.~\ref{sec:gc_sfc_properties}, we develop generating functions for some structural properties of the GC and SFC in projections.

\subsection{Properties of projections}
We consider a random bipartite network of $N$ individual nodes and $M$ group nodes.
Let $p(k)$ and $\tilde{p}(m)$ be degree distributions of individual and group nodes in the bipartite network, respectively.
We define two generating functions that generate $p(k)$ and $\tilde{p}(m)$ as $f_{0}(x)=\sum_k p(k) x^k$ and $g_0(x)=\sum_m \tilde{p}(m) x^m$, respectively.
The derivative of $f_{0}(x)$ ($g_{0}(x)$) gives the generating function of the probability that a randomly chosen edge is incident to an individual (a group) node with degree $k+1$ ($m+1$) as follows:
\begin{align}
	f_1(x)&=\frac{f'_0(x)}{f'_0(1)}=\sum_k \frac{(k+1)p(k+1)}{z_{1}}{x^{k}},\label{eq:f1}\\
	g_1(x)&=\frac{g'_0(x)}{g'_0(1)}=\sum_m \frac{(m+1)\tilde{p}(m+1)}{\tilde{z}_{1}}{x^{m}}.\label{eq:g1}
\end{align}
Here, $z_{1}=\sum_k kp(k)$ ($\tilde{z}_{1}=\sum_m m\tilde{p}(m)$) is the average degree of individual (group) nodes.
The bipartite networks holds the relation $z_{1}N=\tilde{z}_{1}M$.

Next, we consider the projection of a random bipartite network.
The number of cliques and the distribution of the size-$m$ cliques in the projection correspond with the number $M$ of group nodes and group-node degree distribution $\tilde{p}(m)$ in the bipartite network, respectively.
The degree distribution $P(n)$ of the projection is equal to the distribution for the number of the nearest neighboring individual nodes of each individual node in the bipartite network,
and thus, the generating function $G_0(x)$ of the distribution $P(n)$ is given as
\begin{align}
	G_0(x)&=f_0\left(g_1(x)\right).
	\label{eq:G0}
\end{align}
The derivative of $G_{0}(x)$ gives the generating function for the excess degree distribution $Q(n)$ that a randomly chosen edge leads to a node with excess degree $n$, which is the number of edges other than the edge, in the projection as
\begin{equation}
	G_1(x)=\frac{G'_0(x)}{G'_0(1)}=g_2(x)f_1\left(g_1(x)\right),
	\label{eq:G1}
\end{equation}
where $g_2(x)=g'_1(x)/g'_1(1)$.

The global clustering coefficient is defined as $C=3N_{\triangle}/N_{\rm triplet}$, where $N_{\rm triplet}$ and $N_{\triangle}$ denote the numbers of connected triplets and triangles, respectively.
$N_{\rm triplet}$ and $N_{\triangle}$ are given as
\begin{align}
	N_{\rm triplet}&=\frac{1}{2}N\frac{\partial^2}{\partial x^2}G_{0}(x)\bigg|_{x=1}
		=\frac12N\frac{z_1}{\tilde{z}_1}\left(\tilde{z}_{3}+\frac{z_2\tilde{z}_{2}^2}{z_1\tilde{z}_{1}}\right)
	\label{eq:c_num}
\end{align}
and 
\begin{align}
	N_{\triangle}&=\frac{1}{6}M\frac{\partial^3}{\partial x^3}g_{0}(x)\bigg|_{x=1}
	=\frac{1}{6}M\tilde{z}_3,
	\label{eq:c_den}
\end{align}
respectively.
We use the relation $Nz_1=M\tilde{z}_1$ to obtain the global clustering coefficient $C$ of the projection as \cite{newman2001random}
\begin{align}
	C&=\frac{z_1\tilde{z}_1\tilde{z}_3}{z_1\tilde{z}_1\tilde{z}_3+z_2\tilde{z}_{2}^{2}}.
\end{align}

In a manner similar to $G_{0}(x)$, we obtain the generating function $E(x,y)$ of the joint probability $P(n,n')$ such that two ends of a randomly chosen edge in the projection have excess degree $n$ and $n'$ as follows \cite{newman2003social}:
\begin{align}
	E(x,y)&=g_2(xy)f_1\left(g_1(x)\right)f_1\left(g_1(y)\right).
	\label{eq:Exy}
\end{align}
Notably, Eq.~(\ref{eq:G1}) is also obtained from Eq.~(\ref{eq:Exy}) with $y=1$.
The inequality $E(x,y)\neq G_{1}(x)G_{1}(y)$ means that projections can possess degree correlation even though individual and group nodes are randomly connected in bipartite networks, in general.
Using $E(x,y)$ and $G_1(x)$, we can calculate assortativity $r$ as
\begin{equation}
	r=\frac
	{\partial_x\partial_y E(x,y)-\left(\partial_x G_1(x)\right)^2}
	{(x\partial_x)^2 G_1(x)-\left(\partial_x G_1(x)\right)^2}\Bigg|_{x=1,y=1}.
	\label{eq:assortativity}
\end{equation}
Inserting Eqs.~(\ref{eq:G1}) and (\ref{eq:Exy}) into Eq.~(\ref{eq:assortativity}), we obtain \cite{newman2003social}
\begin{align}
	r=\frac{
	1-\cfrac{{\tilde{z}_3}}{{\tilde{z}_2}}
	+\cfrac{{\tilde{z}_4}}{{\tilde{z}_3}}
	}{
	1-\cfrac{{\tilde{z}_3}}{{\tilde{z}_2}}
	+\cfrac{{\tilde{z}_4}}{{\tilde{z}_3}}
	+\cfrac{{z_2} {\tilde{z}_2}}{{z_1} {\tilde{z}_1}}
	\left(1+\cfrac{{\tilde{z}_2}}{{\tilde{z}_3}}+\cfrac{{\tilde{z}_2}^2}{\tilde{z}_1{\tilde{z}_3}}\left(\cfrac{{z_3}}{z_2}-\cfrac{{z_2} }{z_1}\right)\right)}.
	\label{eq:assortativity_moment}
\end{align}
Here $z_n=\sum_k k(k-1)\cdots (k-n+1)p(k)$ and $\tilde{z}_n=\sum_m m(m-1)\cdots (m-n+1)\tilde{p}(m)$.
The assortativity is a non-negative value: $r\geq 0$ (see appendix \ref{sec:assortativity}).

\subsection{Properties of finite components and the giant component}
\label{sec:gc_sfc_properties}
We introduce probability $u$ ($\tilde{u}$) to reach a finite component by an edge outgoing from a group node (an individual node).
The two probabilities are given by self-consistent equations using generating functions Eqs.~(\ref{eq:f1}) and (\ref{eq:g1}):
\begin{align}
	u&=f_1(\tilde{u}), \label{eq:u}\\
	\tilde{u}&=g_1(u)\label{eq:utilde}.
\end{align}
The expanding above near $\epsilon=1-u$ and $\tilde{\epsilon}=1-\tilde{u}$, where $\epsilon~(\tilde{\epsilon})\ll 1$,
we obtain the percolation threshold above which the GC exists as \cite{newman2001random}
\begin{equation}
	\frac{z_2}{z_1}\frac{\tilde{z}_2}{\tilde{z}_1}=1.
	\label{eq:percolation_threshold}
\end{equation}

From the probability $P(n, {\rm SFC})=P(n)u^n$ that a node chosen randomly from the projection has degree $n$ and resides in a finite component, we obtain the generating function $G_{0}(x,{\rm SFC})$ that generates $P(n, {\rm SFC})$:
\begin{equation}
	G_{0}(x,{\rm SFC})=f_0\left(g_1(ux)\right).
\end{equation}
This leads to the generating function $G_{0}(x|{\rm SFC})$ of the degree distribution of the SFC as follows:
 \begin{equation}
	G_{0}(x|{\rm SFC})=\frac{f_0\left(g_1(ux)\right)}{1-S},
	\label{eq:G0fcs}
\end{equation}
where
\begin{equation}
	S=1-f_0\left(g_1(u)\right)
\end{equation}
is the fraction of the GC size.
The relation $G_{0}(x)=G_{0}(x,{\rm SFC})+G_{0}(x,{\rm GC})$ gives related generating functions for the GC as
\begin{equation}
	G_{0}(x,{\rm GC})=f_0\left(g_1(x)\right)-f_0\left(g_1(ux)\right)
	\label{eq:G0gc_joint}
\end{equation}
and
\begin{equation}
	G_{0}(x|{\rm GC})=\frac{f_0\left(g_1(x)\right)-f_0\left(g_1(ux)\right)}{S}.
	\label{eq:G0gc}
\end{equation}
Here $G_{0}(x,{\rm GC})$ is the generating function of the probability that a randomly chosen node has degree $n$ and resides in the GC. $G_{0}(x|{\rm GC})$ is the generating function for the degree distribution of the GC.
In a manner similar to the derivation for Eq.~(\ref{eq:G1}), we obtain the generating functions for the excess degree distribution of the SFC and GC from Eqs.~(\ref{eq:G0fcs}) and (\ref{eq:G0gc}) as
 \begin{align}
	G_1(x|{\rm SFC})&=\frac{g_2(ux)f_1\left(g_1(ux)\right)}{g_2(u)u} 
	\label{eq:g1fc}
\end{align}
and
\begin{align}
	G_1(x|{\rm GC})&=\frac{g_2(x)f_1\left(g_1(x)\right)-ug_2(ux)f_1\left(g_1(ux)\right)}{1-g_2(u)u^2},
	\label{eq:g1gc}
\end{align}
respectively.

From the probability $\tilde{p}(m,{\rm SFC})=\tilde{p}(m)u^m$ that a randomly chosen group node has degree $m$ and resides in a finite component,
we obtain its generating function $g_0(x,{\rm SFC})$ as
\begin{align}
	g_0(x,{\rm SFC})&=g_0(ux).
\end{align}
We replace $G_0(x)$ and $g_0(x)$ in Eqs.~(\ref{eq:c_num}) and (\ref{eq:c_den}) by $G_0(x,{\rm SFC})$ and $g_0(x,{\rm SFC})$, and obtain the number of triplets and triangles in the SFC as
\begin{align}\label{ntriplet}
	N_{\rm triplet}^{\rm SFC}
=\frac{1}{2}Nz_1u^2 \left( ug_1''(u) +(g_1'(u))^2 f_1'(g_1(u)) \right)
\end{align}
and
\begin{align}
	N_{\triangle}^{\rm SFC}=\frac{1}{6}M\tilde{z}_1u^3g_1''(u),
\end{align}
respectively.
The GC is the set difference of the whole network and the SFC,
and thus the numbers for the GC are given by
\begin{align}
	N_{\rm triplet}^{\rm GC}&=N_{\rm triplet}-N_{\rm triplet}^{\rm SFC}
\end{align}
and
\begin{align}
3N_{\triangle}^{\rm GC}&=3N_{\triangle}-3N_{\triangle}^{\rm SFC}.
\end{align}
Thus, the clustering coefficients of the SFC and GC are
\begin{align}
	C^{\rm SFC}=\frac{ug_1''(u)}{ ug_1''(u) +(g_1'(u))^2 f_1'(g_1(u))},
	\label{eq:cglobal_sfc}
\end{align}
and
\begin{align}	
	C^{\rm GC}=\frac{\cfrac{\tilde{z}_3}{\tilde{z}_1}-u^3g_1''(u)}{\cfrac{\tilde{z}_3}{\tilde{z}_1} +\cfrac{z_2\tilde{z}_2^2}{z_1\tilde{z}_1^2} -u^2\bigl(ug_1''(u) + u^2(g_1'(u))^2 f_1'(g_1(u))\bigr)},
	\label{eq:cglobal_gc}
\end{align}
respectively.

Next, we obtain the generating function $E(x,y,{\rm SFC})$ ($E(x,y,{\rm GC})$) for the joint probability $P(n,n',{\rm SFC})$ ($P(n,n',{\rm GC})$) that two ends of a randomly chosen edge have excess degrees $n$ and $n'$ and the edge resides in a finite component (the GC).	
$E(x,y,{\rm SFC})$ and $E(x,y,{\rm GC})$ are given as
\begin{align}
\label{eq:exy_sfc}
	E(x,y,{\rm SFC})&=g_2(uxy)f_1\left(g_1(ux)\right)f_1\left(g_1(uy)\right) 
\end{align}
and
\begin{align}
	E(x,y,{\rm GC})&=g_2(xy)f_1\left(g_1(x)\right)f_1\left(g_1(y)\right)-g_2(uxy)f_1\left(g_1(ux)\right)f_1\left(g_1(uy)\right),
\end{align}
respectively.
We have generating functions $E(x,y|{\rm SFC})$ and $E(x,y|{\rm GC})$ for the joint probabilities of the SFC and GC as
\begin{align}
	E(x,y|{\rm SFC})&=\frac{g_2(uxy)f_1\left(g_1(ux)\right)f_1\left(g_1(uy)\right)}{g_2(u)u^2} \label{eq:exyfc}
\end{align}
and
\begin{align}
	E(x,y|{\rm GC})&=\frac{g_2(xy)f_1\left(g_1(x)\right)f_1\left(g_1(y)\right)-g_2(uxy)f_1\left(g_1(ux)\right)f_1\left(g_1(uy)\right)}{1-g_2(u)u^2}, \label{eq:exygc}
\end{align}
respectively.
We insert Eqs.~(\ref{eq:g1fc}), (\ref{eq:g1gc}), (\ref{eq:exyfc}), and (\ref{eq:exygc}) to Eq.~(\ref{eq:assortativity}) to evaluate the assortativity of the SFC and GC.
Equation (\ref{eq:Exy}) is not equal to both Eqs.~(\ref{eq:exyfc}) and (\ref{eq:exygc}), which implies that the degree correlation of projections is different from their GC and SFC.

\section{Examples}\label{sec:ex}

\subsection{Projections of $z$-uniform bipartite networks}
We consider a random bipartite network whose degree distribution of group nodes is $\tilde{p}(m)=\delta_{mz}$.
Here $\delta_{mz}=1$ for $m=z$ and $\delta_{mz}=0$ otherwise.
We refer to random bipartite networks with $\tilde{p}(m)=\delta_{mz}$ as $z$-uniform bipartite networks.
Projections of $z$-uniform bipartite networks consist of cliques of a single size $z$. Notably, they reduce to ordinary random networks with the degree distribution $p(k)$ when $2$-uniform bipartite networks are chosen.
The generating functions for joint degree distributions of projections of $z$-uniform bipartite networks, their SFC, and their GC are given as
\begin{align}
	E(x,y)&=(xy)^{z-2}f_1(x^{z-1})f_1(y^{z-1}) \label{eq:exy_zurh},\\
	E(x,y|{\rm SFC})&=\frac{(xy)^{z-2}f_1((ux)^{z-1})f_1((uy)^{z-1})}{u^2},
	\label{eq:exyfc_zurh}
\end{align}
and
\begin{align}
	E(x,y|{\rm GC})&=\frac{(xy)^{z-2}f_1(x^{z-1})f_1(y^{z-1})-(uxy)^{z-2}f_1((ux)^{z-1})f_1((uy)^{z-1})}{1-u^z},\label{eq:exygc_zurh}
\end{align}
respectively.
We insert $y=1$ into the above expression to obtain the generating functions for excess degree distributions as
\begin{align}
	G_{1}(x)&=x^{z-2}f_1(x^{z-1}), \label{eq:g1_zurh}\\
	G_{1}(x|{\rm SFC})&=\frac{x^{z-2}{f_1((ux)^{z-1})}}{u}, \label{eq:g1fc_zurh}\\
	G_{1}(x|{\rm GC})&=\frac{x^{z-2}f_1(x^{z-1})-u(ux)^{z-2}f_1((ux)^{z-1})}{1-u^z}.\label{eq:g1gc_zurh}
\end{align}
The equations for $z=2$ coincide with known results for random networks \cite{bialas2008correlations,tishby2018revealing,mizutaka2018disassortativity}.
We find $E(x,y)=G_{1}(x)G_{1}(y)$ and $E(x,y|{\rm SFC})=G_{1}(x|{\rm SFC})G_{1}(y|{\rm SFC})$ from Eqs. (\ref{eq:exy_zurh}), (\ref{eq:exyfc_zurh}), (\ref{eq:g1_zurh}), and (\ref{eq:g1fc_zurh}) without depending on $z$,
which implies that projections of $z$-uniform bipartite networks and their SFC are degree-uncorrelated.
In contrast, the GC is degree correlated [$E(x,y|{\rm GC})\neq G_{1}(x|{\rm GC})G_{1}(y|{\rm GC})$].
From the numerator of Eq.~(\ref{eq:assortativity}) with Eqs.~(\ref{eq:exygc_zurh}) and (\ref{eq:g1gc_zurh}), we obtain
\begin{align}
	\label{eq:rgclq0}
	r^{\rm GC} \propto -\frac{u^z}{(1-u^z)^2}\left.\left( \partial_x G_1(x) - \partial_x G_1(x|{\rm SFC}) \right)^2\right|_{x=1}\leq0.
\end{align}
Equality holds in the above inequality if and only if $u=1$.
The denominator of Eq. (\ref{eq:assortativity}) has a non-negative value, and thus the assortativity of the GC is always negative irrespective of $z$ and $p(k)$.
This implies that when we consider a network with cliques of a single size, its GC is negatively degree correlated for arbitrary individual-node degree distribution $p(k)$.

\begin{figure}
	\begin{center}
		\includegraphics[width=0.45\textwidth]{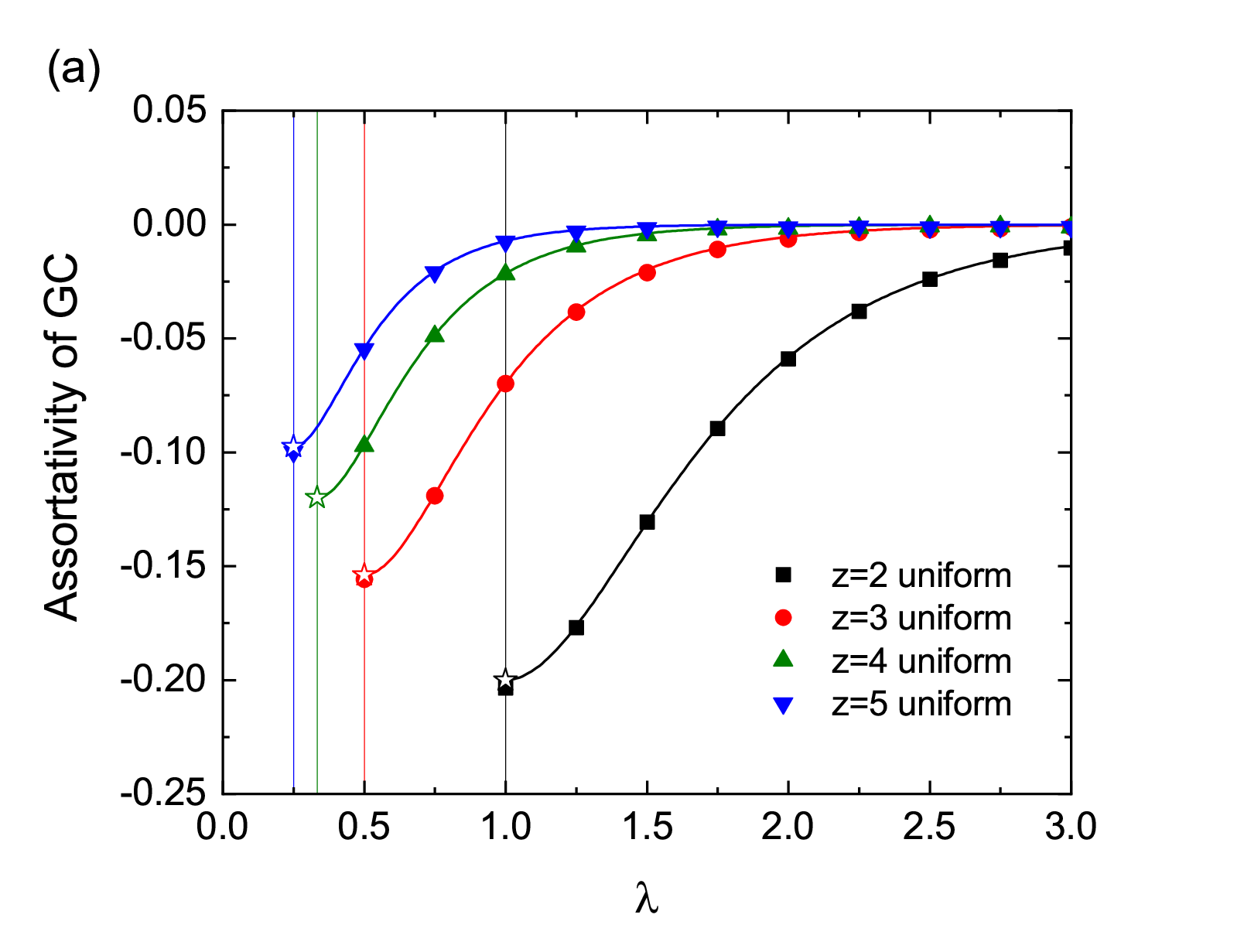}
		\includegraphics[width=0.45\textwidth]{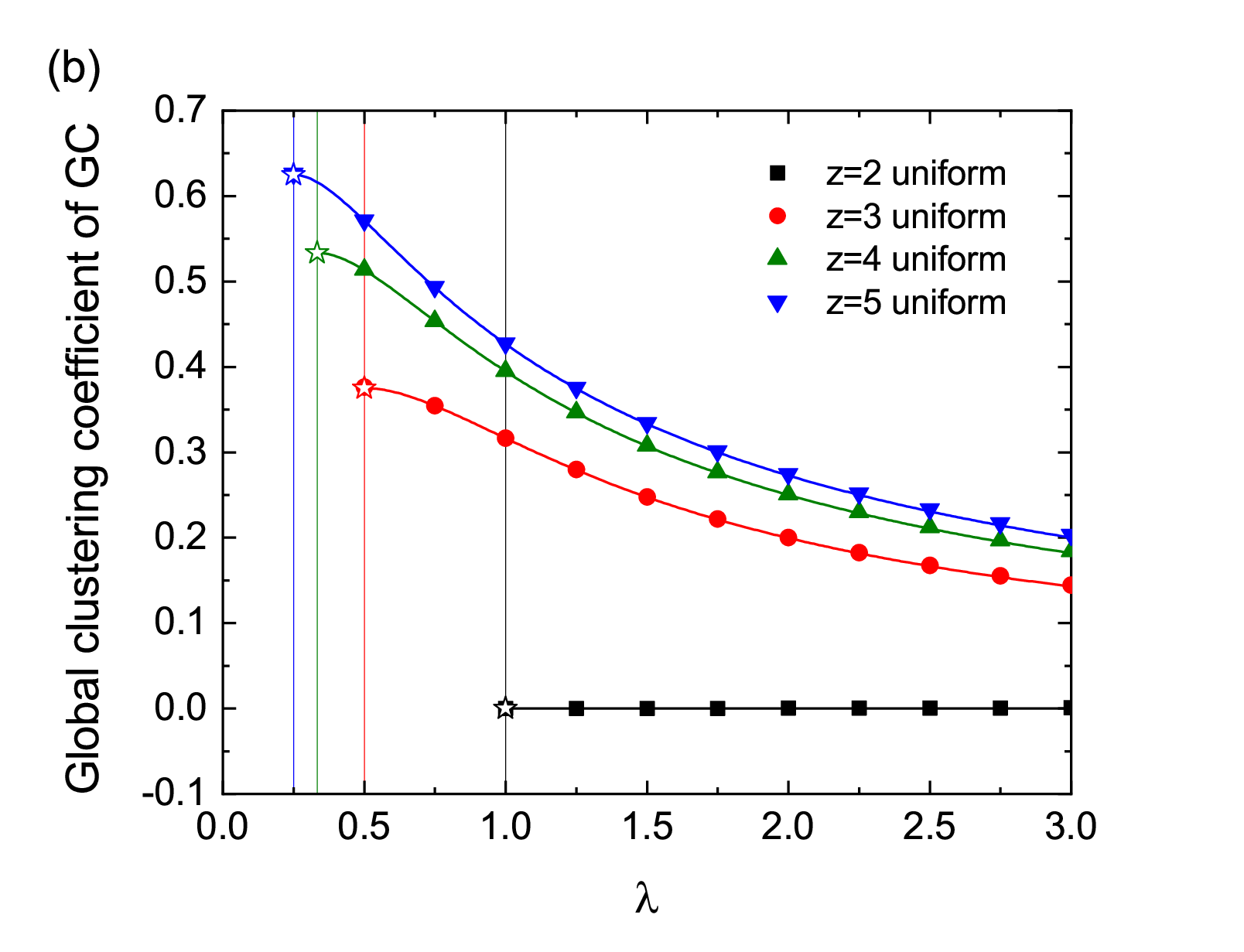}
	\vspace{-10pt}
		\caption{
		\label{fig:fig2}
		Structural properties of the GCs decomposed from the projections of $z$-uniform bipartite networks.
		The vertical axis indicates (a) assortativity and (b) global clustering coefficients.
		The horizontal axis denotes the average of individual-node degrees distributed in a Poisson $p(k)={e^{-\lambda}\lambda^{k}}/{k!}$.
		Each vertical line is the percolation threshold above which the GC exists and a star on the line represents the corresponding (a) $r_{\rm c}^{\rm GC}$ and (b) $C_{\rm c}^{\rm GC}$ given by Eqs.~(\ref{eq:assortativity-atpc-zuni}) and (\ref{eq:cglobal-atpc-zuni}), respectively.
		The solid symbols denote numerical results from Monte Carlo simulations averaged over $1,000$ configurations.
		We set the number of edges between two types of nodes as $W=12,000$.
		The error bars for standard errors on the numerical results are smaller than the data points.
		}
	\end{center}
\end{figure}

We assume that the individual-node degree distribution is Poisson $p(k)={e^{-\lambda}\lambda^{k}}/{k!}$ with the average $\lambda$.
In this situation, the assortativity, global clustering coefficient, and percolation threshold for projections are given as
	\begin{equation}
		r=0,
	\end{equation}
	\begin{equation}
		C=\frac{1}{1+\lambda\frac{z-1}{z-2}},
	\end{equation}
	and 
	\begin{equation}
		\lambda(z-1)=1,
		\label{eq:pc_z-uniform}
	\end{equation}
respectively. From probabilities (\ref{eq:u}) and (\ref{eq:utilde}), we obtain a self-consistent equation
	\begin{equation}
		u=\exp(\lambda(u^{z-1}-1)).
		\label{eq:selfconsistent-zuniform}
	\end{equation}
We evaluate the root $u$ in Eq.~(\ref{eq:selfconsistent-zuniform}) and calculate Eq.~(\ref{eq:assortativity}) with Eqs.~(\ref{eq:exyfc_zurh}) and (\ref{eq:g1fc_zurh}) to obtain the assortativity of the SFC.
Similarly, the assortativity of the GC is derived from Eqs.~(\ref{eq:exygc_zurh}) and (\ref{eq:g1gc_zurh}).
Also, we obtain the global clustering coefficient of the SFC and GC using Eq.~(\ref{eq:cglobal_sfc}) and (\ref{eq:cglobal_gc}), respectively.
If we focus only on the networks at the percolation threshold, the assortativity and global clustering coefficient of the GC can be analyzed. 
We insert $u=1-\epsilon$ into Eq.~(\ref{eq:assortativity}) with Eqs.~(\ref{eq:exygc_zurh}) and (\ref{eq:g1gc_zurh}) and taking $\epsilon\to0$ after some calculations to obtain the assortativity $r_{\rm c}^{\rm GC}$ and global clustering coefficient $C_{\rm c}^{\rm GC}$ of the GC at the percolation threshold (Eq.~(\ref{eq:pc_z-uniform})) as
\begin{equation}
	r_{\rm c}^{\rm GC}=-\frac{z-1}{1+2z(z-1)}
	\label{eq:assortativity-atpc-zuni}
\end{equation}
and
\begin{align}
	C_{\rm c}^{\rm GC}
	=\frac{z(z-2)}{z^2-1},
	\label{eq:cglobal-atpc-zuni}
\end{align}
respectively.
Figure \ref{fig:fig2} (a) and (b) shows the assortativity and global clustering coefficient of the GC as a function of $\lambda$ for $z=2$, $3$, $4$, and $5$.
The lines correspond to the results of Eq.~(\ref{eq:assortativity}) evaluated from generating function methods.
The vertical thin lines represent positions of percolation thresholds, and the symbols (star) are $r_{\rm c}^{\rm GC}$ obtained from Eq.~(\ref{eq:assortativity-atpc-zuni}).
Solid symbols denote numerical results from Monte Carlo simulations.
To generate a bipartite network with $p(k)={e^{-\lambda}\lambda^{k}}/{k!}$ and $\tilde{p}(m)=\delta_{mz}$,
we predetermine the number $W$ of edges between two types of nodes.
Next we prepare $N~(=\lfloor{W/\lambda}\rfloor)$ isolated individual nodes and $M~(=\lfloor{W/z}\rfloor)$ isolated group nodes.
Each group node is connected to randomly chosen $z$ individual nodes.
As shown in Fig.~\ref{fig:fig2}, our analytical treatments are in agreement with the corresponding simulations.
As shown in Fig.~\ref{fig:fig2}(a), the GC is negatively degree-correlated as expected from Eq.~(\ref{eq:rgclq0}).
Notably, the result of Erd\H{o}s--R\'{e}nyi random networks \cite{tishby2018revealing,mizutaka2018disassortativity} is recovered for the case of $z=2$.
Figure~\ref{fig:fig2} (a) shows that the assortativity $r^{\rm GC} (\leq 0)$ increases with $z$.
In general, the correlation coefficient is invariant if two variables are multiplied by a constant.
Given that a $k$-degree individual node in a bipartite network becomes a node with degree $(z-1)k$ in the projection, it is essentially identical for degree correlations in projections of $z$-uniform bipartite networks $(z\geq2)$ as for ordinary random graphs without clustering ($z=2$).
However, the assortativity in different values of $z$ is discrepant because the degrees of nodes involved in GC depends on $k$ and also $z$.
Conversely, the clustering coefficient $C^{\rm GC}$ increases with decreases in $\lambda$ as shown in Fig.~\ref{fig:fig2}~(b).
The average $\lambda$ required to form the GC decreases when the clique size increases, which reduces the number of triplets $N_{\rm triplets}^{\rm GC}$ in the GC.

\subsection{Projections of double Poisson bipartite networks}
Another simple case is a projection from a random bipartite network with $p(k)=e^{-\lambda}\lambda^{k}/{k!}$ and $\tilde{p}(m)=e^{-\tilde{\lambda}}\tilde{\lambda}^{m}/{m!}$.
We refer to such bipartite networks as double Poisson bipartite networks.
The assortativity and clustering coefficient of projections of double Poisson bipartite networks are given as
\begin{equation}
	r=\frac{1}{1+\lambda\tilde{\lambda}+\lambda}
	\label{eq:r_poipoi}
\end{equation}
and
\begin{equation}
	C=\frac{1}{\lambda+1},
\end{equation}
respectively \cite{newman2003social}.
From Eq.~(\ref{eq:percolation_threshold}), the percolation threshold is given as
\begin{equation}
	\lambda\tilde{\lambda}=1.
	\label{eq:dpn_pc_cond}
\end{equation}
Equations (\ref{eq:u}) and (\ref{eq:utilde}) are rewritten as
\begin{align}
	u=\exp(-\lambda (1-\tilde{u})) \label{eq:u_dph}
\end{align}
and
\begin{align}
	\tilde{u} = \exp(-\tilde{\lambda} (1-u)){,}
	\label{eq:utilde_dph}
\end{align}
respectively.

Using Eqs.~(\ref{eq:u_dph}) and (\ref{eq:utilde_dph}),
we determine the following relations among generating functions.
\begin{equation}
	G_{0}(x|{\rm SFC})=G_{0}(x;\lambda^{*}, \tilde{\lambda}^{*}),
	\label{eq:G0fcs_dph2}
\end{equation}
\begin{align}
	G_1(x|{\rm SFC})&=G_{1}(x;\lambda^{*}, \tilde{\lambda}^{*}),
	\label{eq:g1fc_dph2}
\end{align}
and
\begin{align}
	E(x,y|{\rm SFC})&=E(x,y;\lambda^{*}, \tilde{\lambda}^{*}).
	\label{eq:exyfc_dph2}
\end{align}
Here $\lambda^{*}=\tilde{u}\lambda$ and $\tilde{\lambda}^{*}=u\tilde{\lambda}$.
The relations indicate that distributions $P(n|{\rm SFC})$, $Q(n|{\rm SFC})$, and $P(n,n'|{\rm SFC})$ for projections with $\lambda$ and $\tilde{\lambda}$ are identical to distributions $P(n)$, $Q(n)$, and $P(n,n')$ for them with $\lambda^{*}$ and $\tilde{\lambda}^{*}$, respectively.
Hence, the SFC in a supercritical phase ($\lambda\tilde{\lambda}>1$) can be mapped to the whole network with a different parameter set $\lambda^{*}$ and $\tilde{\lambda}^{*}$.
The mapped network is in a subcritical phase ($\lambda^{*}\tilde{\lambda}^{*}<1$) since it does not possess the GC.
This is referred to as a discrete duality relation in studies on random graphs \cite{bollobas2007phase,durrett2007random}.
We can also observe the same relation in projections of $z$-uniform bipartite networks with a Poisson individual degree distribution (not shown).
From the discrete duality relation, the SFC of the projection of double Poisson bipartite networks is positively degree correlated since Eq.~(\ref{eq:r_poipoi}) is positive.

\begin{figure}
	\begin{center}
	\includegraphics[width=0.4\textwidth]{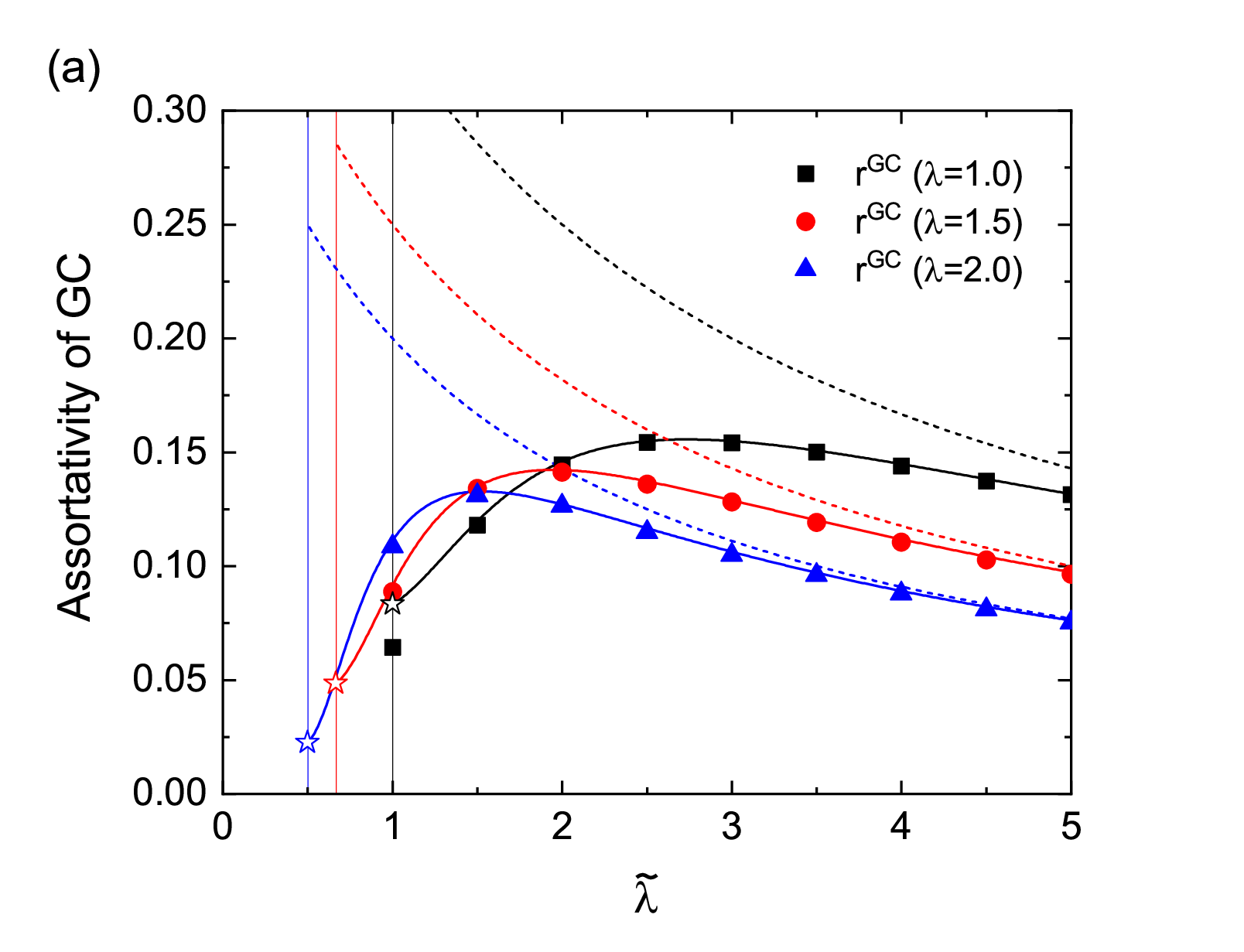}
	\includegraphics[width=0.4\textwidth]{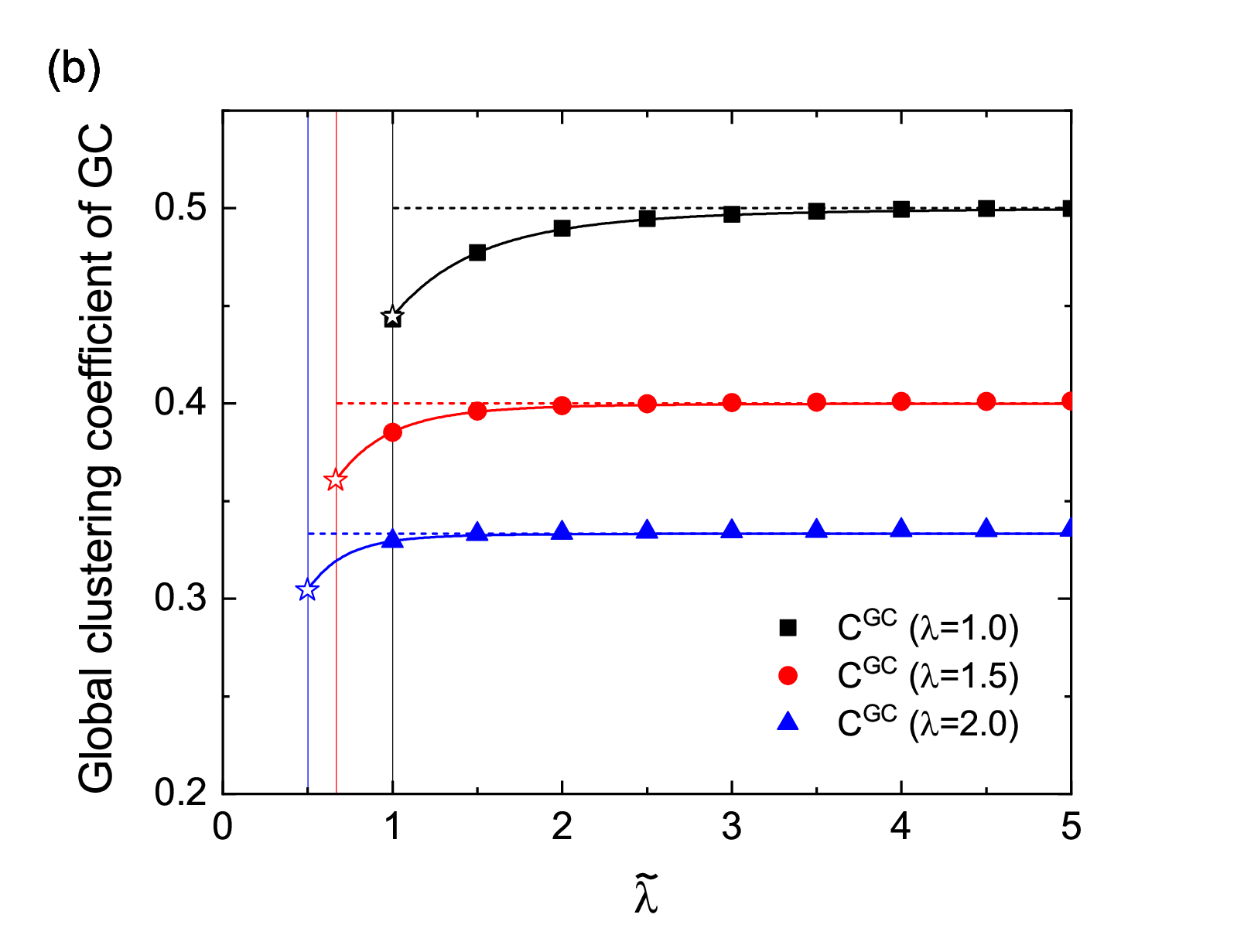}
	\vspace{-10pt}
		\caption{
		\label{fig:fig3}
		Structural properties of projections of double Poisson bipartite networks.
		The lines represent (a) assortativity and (b) global clustering coefficient calculated by the generating function method for the whole network (dashed lines) and GC (solid lines).
		Each vertical line is the percolation threshold and a star on the line represents the corresponding $r_c^{\rm GC}$ and $C_c^{\rm GC}$ given by Eqs. (\ref{eq:rc_double_poisson_bipartite}) and (\ref{eq:cc_double_poisson_bipartite}), respectively.
		The solid symbols represent the numerical results of Monte Carlo simulations averaged over $1,000$ configurations of the GC, which is decomposed from the whole network with $W=12,000$.
		The error bars for standard errors on the numerical results are smaller than the data points.}
	\end{center}
\end{figure}
We determine roots $u$ and $\tilde{u}$ in Eqs.~(\ref{eq:u_dph}) and (\ref{eq:utilde_dph}) under given parameters $\lambda$ and $\tilde{\lambda}$ and calculate Eq.~(\ref{eq:assortativity}) for the SFC and GC to obtain the assortativity of the SFC and GC.
In a manner similar to the derivation of Eq.~(\ref{eq:assortativity-atpc-zuni}), we obtain the assortativity $r_{\rm c}^{\rm GC}$ of the GC at the percolation threshold ($\lambda\tilde{\lambda}=1$) as
\begin{equation}
	r_{\rm c}^{\rm GC}=\frac{\tilde{\lambda}^3+\tilde{\lambda}^2+2\tilde{\lambda}-1}{3\tilde{\lambda}^3+11\tilde{\lambda}^2+17\tilde{\lambda}+5},
	\label{eq:rc_double_poisson_bipartite}
\end{equation}
whose numerator is an increasing function of $\tilde{\lambda}$.
Specifically, $r_{\rm c}^{\rm GC}$ becomes negative for $\tilde{\lambda}\lesssim 0.393$.
We also calculate the global clustering coefficient for the SFC and GC from Eqs.~(\ref{eq:cglobal_sfc}) and (\ref{eq:cglobal_gc}). 
At the percolation threshold, the global clustering coefficient of the GC is
\begin{align}
\label{eq:cc_double_poisson_bipartite}
C_{\rm c}^{\rm GC}(\lambda,\tilde{\lambda})
	=\frac{\tilde{\lambda}^2+3\tilde{\lambda}}{\tilde{\lambda}^2+5\tilde{\lambda}+3}.
\end{align}

\begin{figure}[t]
\begin{center}
\includegraphics[width=1\textwidth]{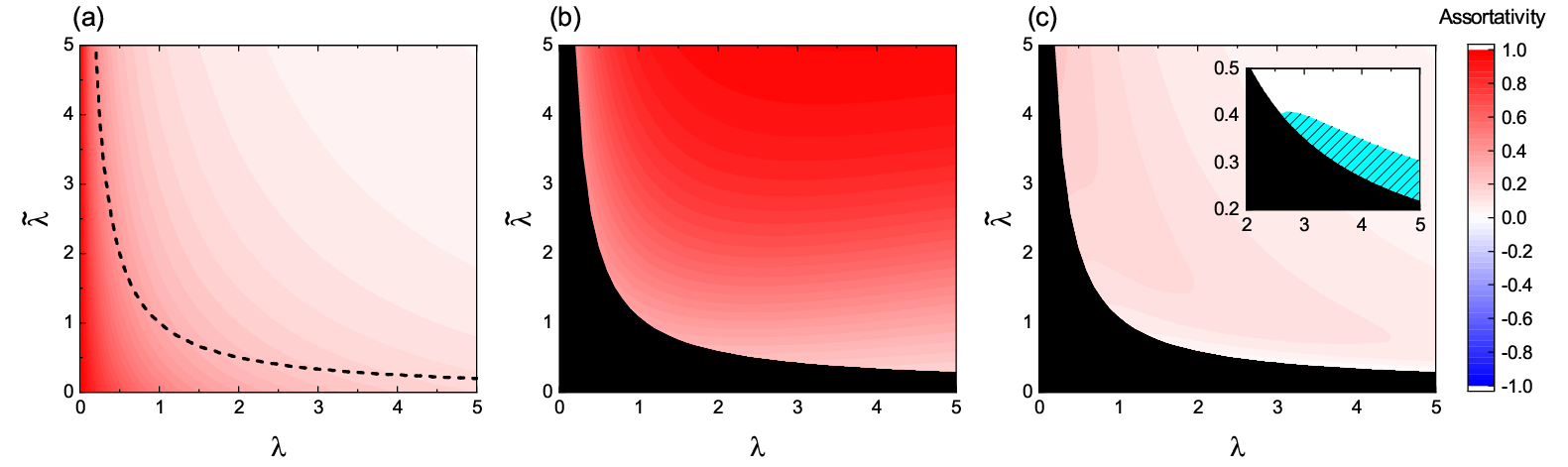}
\caption{
	Assortativities of the whole network (a), SFC (b), and GC (c) of projection of double Poisson bipartite networks.
	The black dashed line in (a) represents the percolation threshold above which there exists the GCs.
	The area filled with black in (b) and (c) denotes subcritical. 
	The hatched area in the inset of (c) is the region where the assortativity of the GC is negative.
	}
\label{fig:fig4}
\end{center}
\end{figure}

Figure \ref{fig:fig3} shows $\lambda$ dependence of assortativity (a) and global clustering coefficient (b) for the whole network (dashed lines) and GC (solid lines) of projection of double Poisson bipartite networks.
Lines are results evaluated from the generating functions.
Symbols are results from Monte Carlo simulations.
In our Monte Carlo simulations, 
to generate double Poisson bipartite networks with $\lambda$ and $\tilde{\lambda}$, 
first, we predetermine the number of edges $W$ between the two types of nodes.
Second, we prepare $N~(=\lfloor{W/\lambda}\rfloor)$ isolated individual nodes and $M~(=\lfloor{W/\tilde{\lambda}}\rfloor)$ isolated group nodes.
Next, we form an edge between a randomly chosen individual node and a randomly chosen group node repeatedly until the number of edges reaches the predetermined $W$.
As shown in Fig.~\ref{fig:fig3}, our analytical treatments are in agreement with the corresponding simulations.
The assortativity $r^{\rm GC}$ and clustering coefficient $C^{\rm GC}$ decrease with $\tilde{\lambda}$ near the percolation threshold.
As $\tilde{\lambda}$ decreases, the size and number of the cliques in the network decreases, which implies that the network structure of projections approaches a locally tree-like structure.
Therefore, the GC tends to lose its positive correlation near the percolation threshold.
Conversely, the degree correlation becomes strongly positive at the SFC consisting of small components with cliques, which is dominant to determine the degree correlation in whole networks and make them positively correlated (See Fig.~\ref{fig:fig4}).

Next we plot color maps of assortativity of the whole network, SFC, and GC in $(\lambda,\tilde{\lambda})$ plane at the panel (a), (b), and (c), respectively of Fig.~\ref{fig:fig4}.
The dashed line in the panel (a) represents the percolation threshold given as Eq.~(\ref{eq:dpn_pc_cond}).
The black region in the panel (b) and (c) is subcritical where the GC does not exist.
We confirm from the panel (a) that the assortativity of the whole network is a decreasing function of $\lambda$ and $\tilde{\lambda}$ as expected by Eq.~(\ref{eq:r_poipoi}).
The SFC exhibits a strong and positive degree correlation (Fig.~\ref{fig:fig4} (b)). Isolated cliques in the SFC contribute to a positive degree correlation.
Figure~\ref{fig:fig4} (c) shows that the degree correlation of GC is weak.
The assortativity $r^{\rm GC}$ is positive for almost the entire panel (c).
The inset of (c) shows that $r^{\rm GC}$ near the percolation threshold can be negative with small $\tilde{\lambda}$ 
[e.g. $r^{\rm GC}\approx-0.003$ at the point of $\tilde{\lambda}=0.30$ and $\lambda=5.00$ and see also Eq.~(\ref{eq:rc_double_poisson_bipartite})].

\section{Conclusion and Discussion}\label{sec:end}
In the study, we examined statistical properties of projections of random bipartite networks characterized by the individual node degree distribution $p(k)$ and group node degree distribution $\tilde{p}(m)$.
The projections can be decomposed into two subgraphs: the giant component (GC) and set of finite components (SFC).
We presented a method using generating functions to evaluate the structural properties of both subgraphs and derived the assortativity and global clustering coefficient of the GC and SFC.
We treated two examples in this study.
The first example is $z$-uniform bipartite networks, and the other is double Poisson bipartite networks.
With the exception of the $2$-uniform bipartite network, projection, GC, and SFC of the examples have non-zero clustering coefficients.
Conversely, assortativity varies based on the subgraph and bipartite network model.
The assortativity values of the networks are summarized in Table~I.
For a $z$-uniform bipartite network, we analytically demonstrated that the projection and SFC have no degree correlation. Simultaneously, the GC is negatively correlated irrespective of the value of $z$ and the distribution $p(k)$ unless the whole network corresponds to the GC.
In contrast, for a double Poisson bipartite network, the projection and SFC display strong and positive degree correlations.
In addition, the GC exhibits a weak positive correlation in a wide range of parameter sets ($\lambda$, $\tilde{\lambda}$) and a weak negative correlation in a narrow region.
Here, $\lambda$ and $\tilde{\lambda}$ correspond to the average of $p(k)$ and $\tilde{p}(m)$, respectively.
In the region where negative correlation is observed, $\tilde{\lambda}$ is small and most of the cliques involved in the GC are of small size, including a simple edge, resulting in a tree-like structure.
\begin{table}[tbtb]
\caption{Assortativity of projections in the thermodynamic limit.}
\label{table:1}
\centering
\begin{tabular}{cccccc}
	\hline \hline
	~~~~~Bipartite network~~~~~& ~~~~$p(k)$~~~~ & ~~~~$\tilde{p}(m)$~~~~ & ~~~~whole~~~~ & ~~~~SFC~~~~ & ~~~~GC~~~~ \\
	\hline
	$z$-uniform
	& arbitrary	& $\delta_{mz}$
	& $0$	& $0$	& $\leq0$\\
	double Poisson
	& Poisson	& Poisson
	& $>0$ & $>0$	& $\gtrsim0$ \\
	\hline \hline
\end{tabular}
\end{table}

\begin{figure}[t]
\begin{center}
	\includegraphics[width=0.4\textwidth]{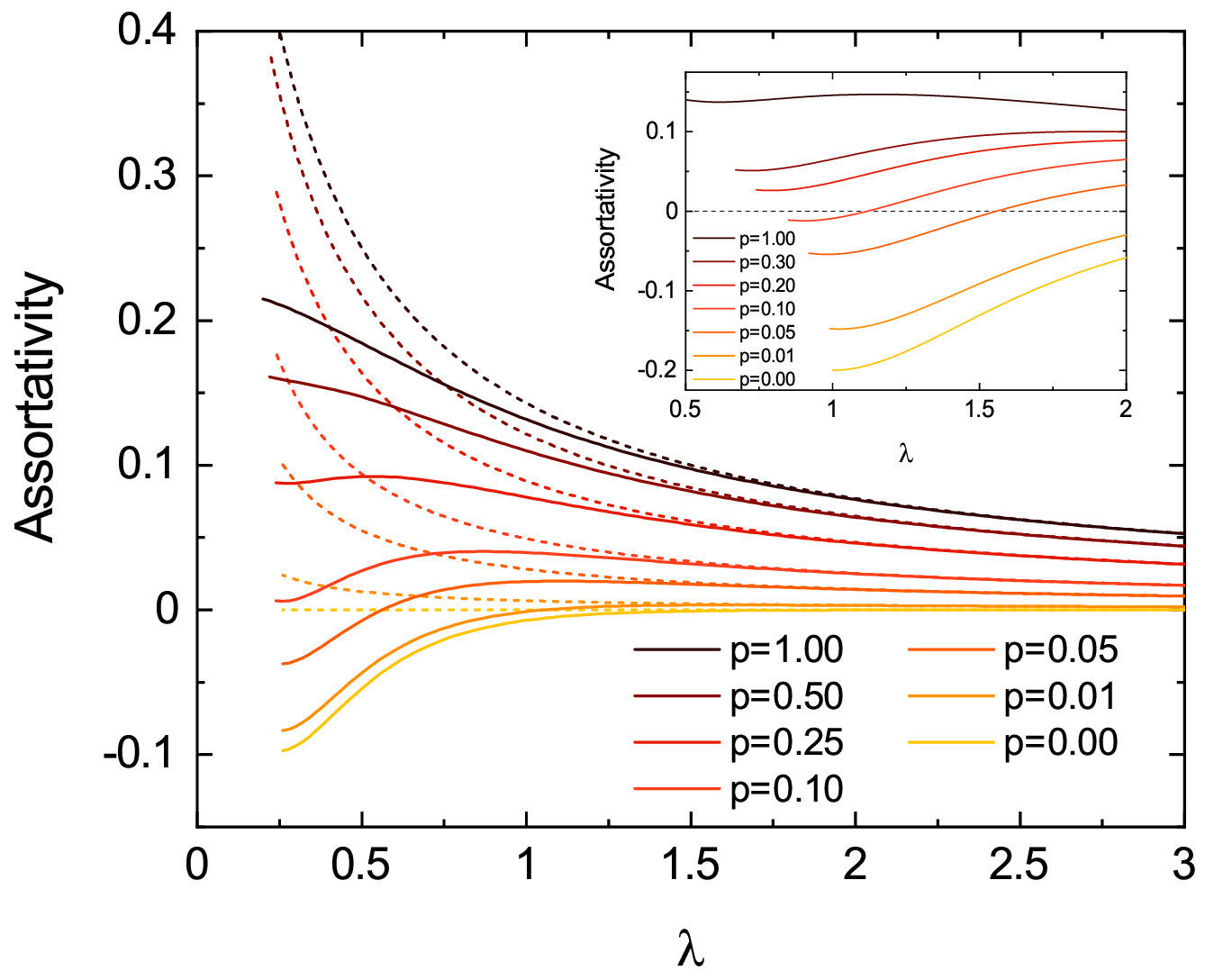}
	\vspace{-10pt}
	\caption{ \label{fig:fig5}
	Comparison of assortativity of the GC (solid lines) and whole network (dashed lines) for $z=5$.
	The inset is for the GC of the projection when $z=2$.
	}
\end{center}
\end{figure}

We consider the same $p(k)$ (the Poisson distribution) for both examples in Figs.~\ref{fig:fig2} and \ref{fig:fig4}.
Thus, the difference in degree correlations arises from $\tilde{p}(m)$ that corresponds to the clique size distribution in projections.
To confirm the effect of the fluctuation of the clique sizes, we consider the following bipartite network model, which connects $z$-uniform bipartite networks and double Poisson bipartite ones continuously.
First, we prepare a bipartite network with $W$ edges whose degree distributions obey $p(k)={e^{-\lambda}\lambda^{k}}/{k!}$ and $\tilde{p}(m)=\delta_{mz}$.
Next, a randomly selected edge is removed from the group node side and connected to a randomly selected group node.
The randomization repeats $W'$ times.
Generated bipartite networks are identical to the $z$-uniform bipartite networks (double Poisson bipartite networks) if the fraction $p=W'/W$ of edge replaces is zero (one).
The variance of $\tilde{p}(m)$ is given as $\sigma_m^2=pz(2-p)$ when the number of group nodes is large enough (see appendix \ref{sec:variance}), and thus the fraction $p$ tunes the clique size fluctuation.
Figure~\ref{fig:fig5} shows the assortativity for $z=5$ as a function of the average $\lambda$ of $p(k)$ for projections with different values $p$.
The solid and dashed lines represent the result for the GC and whole networks, respectively.
For $p<0.05$, we observe that the assortativity of GC decreases with $\lambda$ and takes a negative value, while for $p>0.05$, it increases with decreases in $\lambda$.
The inset of Fig.~\ref{fig:fig5} shows the results for $z=2$. 
For $p=0$, the projections coincide with ordinary random graphs that must exhibit the strongest negative degree correlation in the present setting.
Even in this case, the GC does not show a negative degree correlation for $p\gtrapprox 0.1$.
Thus, a small amount of fluctuation (small $p$) makes the degree correlation of GC positive, which implies that projections of most bipartite networks display a positive degree correlation including their GCs.

In the study, the relation between the SFC in a supercritical phase possessing the GC and the projection in a subcritical phase for a double Poisson bipartite network is determined.
The relation claims that the SFC in a supercritical phase can be mapped to the projection in a subcritical phase; i.e., they have the same structure.
Thus, given that the assortativity $r$ for whole networks is non-negative [Eq.~(\ref{eq:assortativity_moment})], the assortativity $r^{\rm SFC}$ is also non-negative.
Whether it holds for the general case is left for future studies.
Clarifying this point is essential to understand bipartite networks and their projections.

\acknowledgments
Y.\ F.\ was supported by JSPS KAKENHI Grant Number 21K21302 and 23K13010.
S.\ M.\ was supported by JSPS KAKENHI Grant Number 21K13853.

\appendix
\section{Assortativity for projections}
\label{sec:assortativity}
Equation~(\ref{eq:assortativity_moment}) can be rewritten as
\begin{align}
	r
	&=\frac{\tilde{\beta}_2}
	{\tilde{\beta}_2+\cfrac{{z_2}{\tilde{z}_2}^2}{{z_1} {\tilde{z}_1}^2}
	\left(\tilde{\beta}_1+\cfrac{{\tilde{z}_2}^2}{z_1z_2}\beta_1\right)},
\end{align}
where
\begin{align}
	\beta_n=\cfrac12\sum_{k_1,k_2} (k_1-k_2)^2
	\left[p(k_1)\prod_{l=0}^{n-1}(k_1-l)\right]\left[p(k_2)\prod_{l=0}^{n-1}(k_2-l)\right]\geq0
\end{align}
\begin{align}	
	\tilde{\beta}_n=\cfrac12\sum_{m_1,m_2} (m_1-m_2)^2
	\left[\tilde{p}(m_1)\prod_{l=0}^{n-1}(m_1-l)\right]\left[\tilde{p}(m_2)\prod_{l=0}^{n-1}(m_2-l)\right]\geq0
\end{align}
are non-negative values.
Thus, the assortativity of a projection of a random bipartite network is always non negative.
If the clique size fluctuation exists, the value of Eq.~(\ref{eq:assortativity_moment}) is always positive: $r>0$.

\section{Variance of clique size fluctuation}
\label{sec:variance}
Here we derive the variance of group-node degrees using the generating function method.
The probability that $m$ edges are connected to a group node after random trimming of $W'$ edges is
\begin{align}
	P_{\rm trim}(m)=\binom{z}{m}p^{z-m}(1-p)^{m}.
\end{align}
The probability that $m$ edges are added to a group node is
\begin{align}
	P_{\rm add}(m)=\binom{W'}{m}{q}^{m}(1-q)^{W'-m},
\end{align}
where $q=1/M$.
The generating function of the probabilities are
\begin{align}
	G_{\rm trim}(x)&=(p+(1-p)x)^z,\\
	G_{\rm add}(x)&=(qx+1-q)^{W'}.
\end{align}
The degree distribution of group nodes $\tilde{p}(m)$ after the trimming and adding $W'$ edges can be represented as the convolution of $P_{\rm trim}(m)$ and $P_{\rm add}(m)$:
\begin{align}
	\tilde{p}(m)=\sum_{m'}P_{\rm trim}(m')P_{\rm add}(m-m').
\end{align}
The generating function of a convolution is the product of the generating functions of the indices of the convolution, and thus,
the generating function of the $\tilde{p}(m)$ is
\begin{align}
	g_0(x)&=\left(p+(1-p)x\right)^z \left(qx+1-q\right)^{W'}.
\end{align}
The variance of group-node degrees is
\begin{align}
	\sigma_m^2=pz(2-p-q).
\end{align}
The value is an increasing function of the edge randomization probability $p$.

\printindex
\end{document}